\documentclass{aa}  

\usepackage{graphicx}
\usepackage[dvipsnames]{xcolor}

\usepackage{txfonts}
\usepackage[amsmath,thref,hyperref]{ntheorem}
\usepackage[colorlinks=true,
    linkcolor=blue,
    citecolor=blue,
    filecolor=magenta,      
    urlcolor=cyan]{hyperref}
\usepackage{orcidlink}

\begin{document}

   \title{On the behaviour of eccentric sub-pc massive black hole binaries embedded in massive discs}
   \titlerunning{Eccentric MBHBs}
   \authorrunning{A. Franchini et al.}

   \author{
          Alessia Franchini\orcidlink{0000-0002-8400-0969}
          \inst{1,}
          \inst{2,}          \inst{3}\fnmsep\thanks{alessia.franchini@uzh.ch}
          \and
          Alessandra Prato
          \inst{2}
          \and
          Cristiano Longarini\orcidlink{0000-0003-4663-0318}
          \inst{4,5}
          \and
          Alberto Sesana\orcidlink{0000-0003-4961-1606}
          \inst{2,}
          \inst{3}
    }

   \institute{
            Universität Zürich, Institut für Astrophysik, Winterthurerstrasse 190, CH-8057 Zürich, Switzerland
        \and
            Dipartimento di Fisica ``G. Occhialini'', Universit\`a degli Studi di Milano-Bicocca, Piazza della Scienza 3, I-20126 Milano, Italy
        \and
            INFN, Sezione di Milano-Bicocca, Piazza della Scienza 3, I-20126 Milano, Italy
        \and
            Institute of Astronomy, University of Cambridge, Madingley Rd, Cambridge, CB3 0HA, United Kingdom
        \and
            Dipartimento di Fisica, Universit\`a degli Studi di Milano, Via Celoria 16, 20133 Milano, Italy
        }

   \date{Received xxx / Accepted xxx}

 
  \abstract
   {Using the 3D smoothed particle hydrodynamics code {\sc phantom}, we investigate the evolution of the orbital properties of massive black hole binaries embedded in massive discs where gravitational instabilities (GIs) triggered by the disc self-gravity are the only source of angular momentum transport. In particular, we investigate the evolution of binaries with different initial eccentricities $e_0=0.05,\,0.5,\,0.8$ and mass ratios $q=0.1,\,0.3,\,0.9$.
    Our simulations suggest that there might not be a unique value of critical eccentricity. 
    We find initially more eccentric binaries to tend to higher asymptotic eccentricity values than more circular ones. This implies that there is a range of critical eccentricity values, that depends on the initial condition of the system. In particular, we find the width of this range to be narrower for more unequal binaries.
    We furthermore measure preferential accretion onto our binaries, finding more accretion onto the primary only for mass ratio $q=0.3$ and eccentricity $e=0.8$. We discuss how this might have implications for the amplitude of the gravitational wave background detected by Pulsar Timing Arrays (PTA) experiments. We finally measure the corresponding value of the viscosity parameter $\alpha$ in our simulations and discuss how this depends on the binary properties.}


   \maketitle
%

\section{Introduction}
Observational evidence suggests that massive black holes (MBHs) inhabit the nuclei of (virtually all) massive galaxies \citep[][and references therein]{Kormendy2013}. When two galaxies merge, the MBHs hosted in their nuclei migrate to the center of the merger remnant primarily due to dynamical friction against the background of stars and gas \citep{Chandrasekhar1943}. At parsec separations the two black holes bind into a massive black hole binary (MBHB). At this point, dynamical friction becomes inefficient and further evolution of the binary requires a physical mechanism able to extract its energy and angular momentum. The main mechanisms proposed in the literature are three-body scattering of stars intersecting the binary orbit \citep{Quinlan1996,sesana2007,Bortolas2021} or the interaction with a circumbinary gaseous disc \citep{Dotti2007,mayer2007b,Lodato2009,Dotti2012}. 
Since both galaxies might contain large amounts of gas, it is expected that this will sink to the centre of the newly formed galaxy and form a circumbinary accretion disc \citep{begelman1980,escala2005,Cuadra2009}. 

The understanding of the disc-binary coupled dynamics is crucial since at sub-pc scales the binary is too wide for GWs emission to take over driving the binary to merge.
The presence of such a disc around a MBHB might facilitate its merger and potentially give rise to observational signatures in the form of electromagnetic (EM) signals \citep{mm2001,ArmitageNatarajan2002,Lodato2009}. 

Very early numerical simulations \citep{artymowicz1994,artymowicz1996} investigated the interaction of a binary with its gaseous circumbinary disc finding that only a small amount of material is able to enter the cavity carved by the binary and to accrete onto the binary components. The main finding of these works is that the binary semi-major axis decreases with time owing to the interaction with the disc.
This picture has been recently challenged by a few works \citep{Munoz2019,Duffell2019,Munoz2020} employing 2D (and one 3D, see \citealt{Moody2019}) static or moving-mesh grid numerical simulations with fixed binary orbits. In particular, these studies found that the secular angular momentum transfer onto the binary is strongly positive within the range of binary and disc parameters explored. More recently \cite{tiede2020} found, using the same numerical techniques, that the sign of the torque exerted by the disc onto the binary depends on the disc temperature, i.e. on its aspect ratio $H/R$, for locally isothermal discs. In particular, they found discs with $H/R \lesssim 0.04$ to shrink the binary. 
Using 3D smoothed particle hydrodynamics (SPH) simulations of locally isothermal discs, \cite{heathnixon2020} found instead the threshold value for binary expansion to be $H/R\simeq 0.2$.
However, \cite{Franchini2022} found this process not to be simply regulated by the disc temperature but also by the disc viscosity. 

All these previous works did not include the disc self-gravity which is negligible for relatively light ($M<10^7$M$_\odot$) compact ($a<1$ mpc) binaries \citep[see discussion in][]{franchini2021}. In the regime where the disc self-gravity cannot be neglected, and the disc temperature changes with time, all previous works found binary shrinking as a result of the interaction with massive discs regardless of the disc temperature \citep{Cuadra2009,roedig2012,franchini2021}.

The eccentricity evolution of binaries embedded in massive self-gravitating discs was first investigated by \cite{roedig2011}, finding that binaries with mass ratio $q=M_2/M_1=0.3$ evolve towards a critical eccentricity in the range $0.6-0.8$.
The important consequence of the existence of some $e_{\rm crit}$ is the detectability of a significant residual eccentricity by the Laser Interferometer Space Antenna (LISA) \citep{2023LRR....26....2A}.  
The excitation of MBHB eccentricity in gaseous discs has a significant impact also on Pulsar Timing Arrays (PTA) sources. Many of these are at the evolutionary stage in which the environment still plays a role in their dynamical evolution and they are therefore expected to be significantly eccentric. This must be appropriately taken into account when developing dedicated data analysis algorithms.
The investigation of these systems is timely to aid the interpretation of the GW signal recently detected by pulsar timing array (PTA) collaborations around the globe; namely the European PTA \citep[EPTA,][]{EPTAI2023,EPTAII2023,EPTAIII2023,EPTAIV2023,EPTAV2023,Smarra2023}, NANOGrav \citep{nanograv2023,2023ApJ...951L...8A,2023ApJ...951L...9A,2023ApJ...951L..10A,2023ApJ...951L..11A},  the Parkes PTA \citep[PPTA,][]{ppta2023} and the Chinese PTA \citep[CPTA,][]{2023RAA....23g5024X}.

A few recent works studied the evolution of the binary eccentricity driven by the presence of the circumbinary disc \citep{Zrake2021,Siwek2023}. They find that binaries with large, close to equal, mass ratio all evolve towards $e\sim 0.5$, regardless of their initial conditions.
These studies however employ 2D fixed binary orbit numerical simulations and neglect the disc self-gravity, therefore focusing on a specific region of the parameter space where the disc self-gravity can be neglected \citep{franchini2021}.

In this work we consider the dynamics of MBHBs in massive discs governed by self gravity, expanding on the work of \citep{roedig2011}. In particular, we further explore the parameter space investigating the effect that the binary mass ratio has in determining the critical eccentricity value.
The paper is organized as follows. In Section \ref{sec:numerical} we outline the numerical setup. We present our results in Section \ref{sec:results} and draw our conclusions in Section \ref{sec:conclusions}.

\section{Numerical Setup}
\label{sec:numerical}

The live binary \citep{Franchini2022,Franchini2023} is composed by two sink particles \citep{bate1995} with initial separation $a=1$ and total mass equal $M=M_1+M_2=1$ in code units. The initial binary orbital period is therefore $P_{\rm b}=2\pi$.
We take the two sinks to have the same accretion radius $r_{\rm sink}=0.03a$. Particles inside this radius are accreted onto the respective sink particle without any further check on their angular momentum or energy.
We run several numerical simulations varying the initial binary and disc parameters. 
The initial value of binary eccentricity ranges in the interval $0.05-0.8$, i.e. from almost circular to very eccentric binaries while the binary mass ratio was assumed to be $q=0.1, 0.3, 0.9$.

The disc initially extends from $R=2a$ to $R=5a$, where $a$ is the binary separation, has mass $M_{\rm d}=0.2M$ and is composed of $N=10^6$ equal-mass gas particles distributed according to a power law profile $\Sigma \propto R^{-1}$.

We take the disc to be corotating and aligned with the binary orbital plane. The disc equation of state is adiabatic with $\gamma=5/3$ and we take the \cite{gammie2001} cooling function, with a cooling factor $\beta= t_{\rm cool}/t_{\rm dyn}=10$ in order for the disc to be sufficiently unstable to form large spirals but not enough to fragment, since we are not interested in the latter.
We assume the disc initial aspect ratio to be $H/R=0.2$ which gives an initial value of the Toomre parameter $Q=1.50$ at the disc outer edge. 
The transport of angular momentum throughout the disc is regulated by Gravitational Instabilities (GIs) \citep{lodato2007sg,meru2012}.

We apply artificial viscosity only to approaching particles in order to be able to resolve shocks, using
the \cite{cullen2010} switch. The shock capturing dissipation terms are included in the code according to the approach outlined in \cite{monaghan1997} and consist of a linear term $\alpha_{\rm AV}$, controlled by the switch, and a quadratic term $\beta_{\rm AV}$, which essentially prevents particle interpenetration. Since we want the transport of angular momentum induced by gravitational instabilities to dominate, we minimize the numerical dissipation introduced by the numerical viscosity by setting $\alpha^{\rm min}_{\rm AV}=0,\,\alpha^{\rm max}_{\rm AV}=1$
and $\beta_{\rm AV} = 2.0$ \citep{meru2012}.

We run all the simulations for $\sim 1000$ orbits, taking a snapshot every $0.1\,P_{\rm b}$. Note that $1000$ orbits corresponds to $\sim 9$ cooling times at the disc outer edge. We can therefore assume the disc to have reached a stable gravitoturbulent configuration.

\section{Results}
\label{sec:results}

\begin{table}
    \centering
    \begin{tabular}{c|c|c|c|}
    \hline
    $q$ & $e_0$ & $M_{\rm disc}$ & H/R\\
    \hline
    \hline
     $0.1$ & $0.05, 0.5, 0.8$ & 0.2M & 0.2\\
     \hline
     $0.3$ & $0.05, 0.5, 0.8$ & 0.2M & 0.2\\
     \hline
     $0.9$ & $0.05, 0.5, 0.8$ & 0.2M & 0.2\\
     \hline
    \end{tabular}
    \caption{Simulations parameters. First and second column contain the binary initial mass ratio and eccentricity while the third and forth column report the initial disc mass and   initial aspect ratio respectively.}
    \label{tab:sim_params}
\end{table}

We explore the parameter space by running a set of simulations varying the initial binary eccentricity and mass ratio. The values are reported in Table \ref{tab:sim_params}.
Note that we did not vary either the disc mass or aspect ratio as these control the stability of the disc to GIs and their effect has already been investigated in \cite{franchini2021}.

\begin{figure*}
    \includegraphics[width=\textwidth]{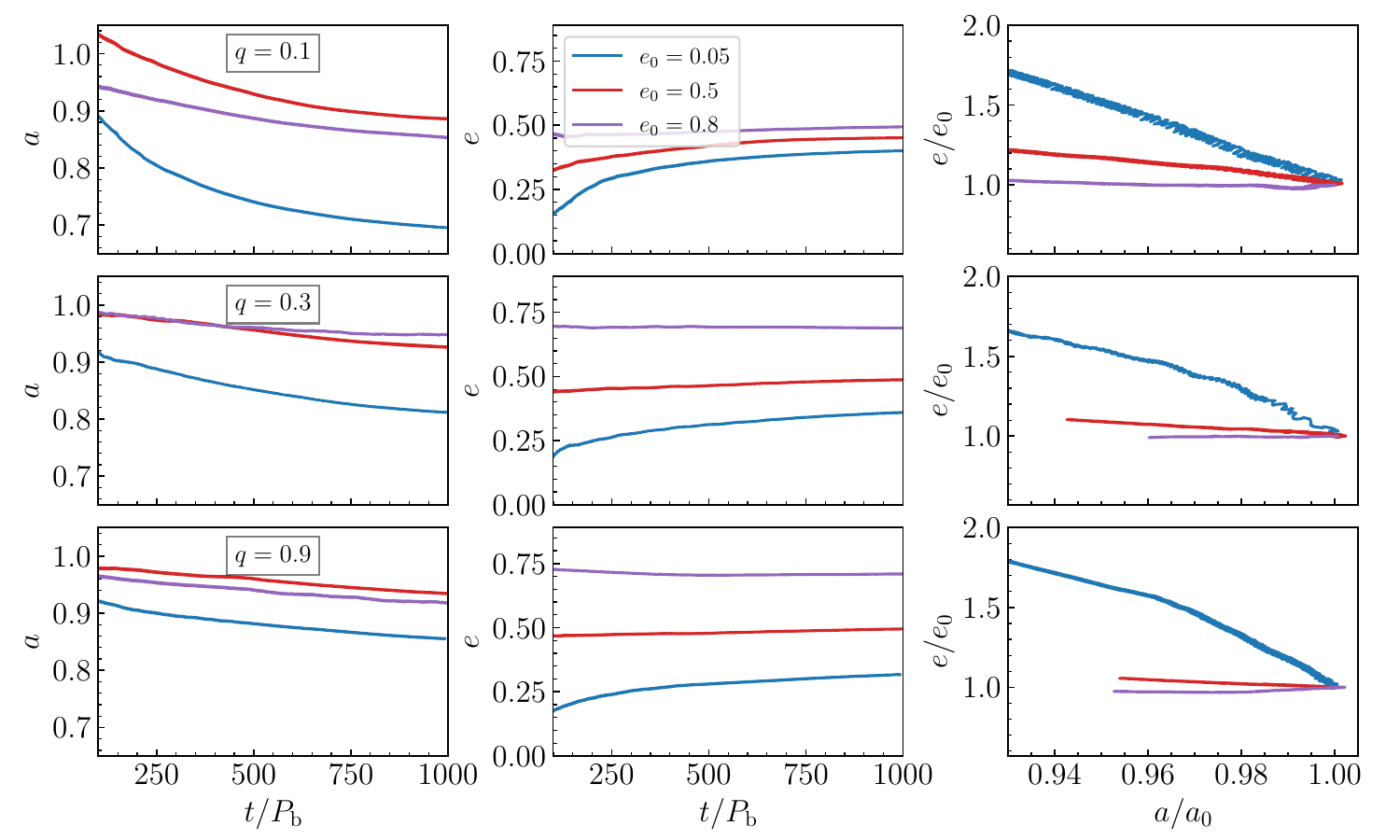}
    \caption{Evolution of the binary semi-major axis and eccentricity for initial mass ratio $q=0.1$ (upper panels), $q=0.3$ (middle panels) and $q=0.9$ (bottom panels). The blue, red and purple lines show the simulations with $e_0=0.05,\,0.5,\,0.8$ respectively. Note that here we do not show the transient phase occurring within the first 100 orbits. Therefore $e_0$ and $a_0$ in the last column are the values of eccentricity and semi-major axis after 100 orbits.}
    \label{fig:aeevo}
\end{figure*}

\subsection{Critical eccentricity}

The increase of the eccentricity caused by the interaction of the binary with an external disc is a known fact for very unequal binaries where the non-axisymmetric potential perturbations are small \cite{goldreich1980}. 
In the high mass-ratio limit, the transfer of angular momentum from the binary to the disc occurs secularly, through torques excited in the disc by the binary at discrete Lindblad and co-rotation resonances. Damping or growth of eccentricity thus depends on the relative importance of these opposing torques (and so on how fluid elements are distributed in the disc). Lindblad resonances are known to be responsible for opening a gap in the disc. As a consequence of disc clearance, co-rotation and inner Lindblad resonances are reduced in power. \cite{Goldreich2003} showed that only the outer Lindblad resonances, remaining after gap opening, cause the increase of the eccentricity for initially low eccentric binaries.

The existence of a critical eccentricity of equilibrium ,$e_{\rm crit}$, in comparable mass binaries has been extensively discussed in \cite{roedig2011}. Their main argument is essentially linked to the gravitational pull of the disc onto the binary.
A binary with small initial eccentricity can become more eccentric because of the larger deceleration experienced by the secondary MBH near apocentre with respect to pericentre. The longer time spent near the apocentre and the larger over-density excited in the disc by the MBH gravitational pull due to its immediate proximity lead to a net deceleration of the MBH that causes the eccentricity increase. This process continues as long as the secondary MBH has a larger angular velocity at its apocentre than the fluid elements in the disc. When this reverses, the density wake excited by the MBH moves ahead imparting to the MBH a net tangential acceleration that tends to increase the angular momentum content of the binary, decreasing its eccentricity.
The efficiency of this mechanism has not been, however, extensively investigated. In particular, \cite{roedig2011} considered only systems with $q=1/3$ and they extrapolated 
their results to other mass ratios based on analytical arguments. Furthermore, it is unclear whether, for a given $q$, a well defined value of $e_{\rm crit}$ exists, independent on the initial eccentricity of the system. We investigate these points in the following discussion.

The evolution of the binary orbital parameters is shown in Figure \ref{fig:aeevo}. The first column shows the semi-major axis and the second shows the eccentricity. The last column shows the evolution of the eccentricity as a function of the semi-major axis, both normalized to their value at orbit 100.
Note that we do not show the initial transient phase that the binaries in our simulation experience due to the initial, out of equilibrium, disc configuration.

As a general trend, we find that the  binary separation always decreases over time, at a rate that is only mildly affected by the binary eccentricity. Conversely, for any given mass ratio, the eccentricity evolution slows down significantly with time and depends on $e_0$, with more eccentric systems showing slower evolution, indicative of saturation.

We can take a closer look at the comparable mass binary $q=0.9$, shown in the bottom panels of Figure \ref{fig:aeevo}. The semi-major axis evolution rate (left panel) is essentially equivalent for all values of $e_0$, and roughly constant in time. Conversely, the eccentricity evolution (central panel) is very different in the three cases. The binary with $e_0=0.05$ shows a significant eccentricity growth, reaching $e\approx 0.35$ after 1000 orbits. The growth then seems to continue, although with a slower rate. 
At the opposite end, the initially eccentric binary, $e_0=0.8$, experience a slow eccentricity dumping down to $\approx 0.7$. Note that after few hundred orbits, $e$ essentially saturates and stays constant. Finally, the medium-eccentricity system, $e_0=0.5$, only displays a marginal evolution, which is again indicative of saturation.

\begin{figure}
    \includegraphics[width=\columnwidth]{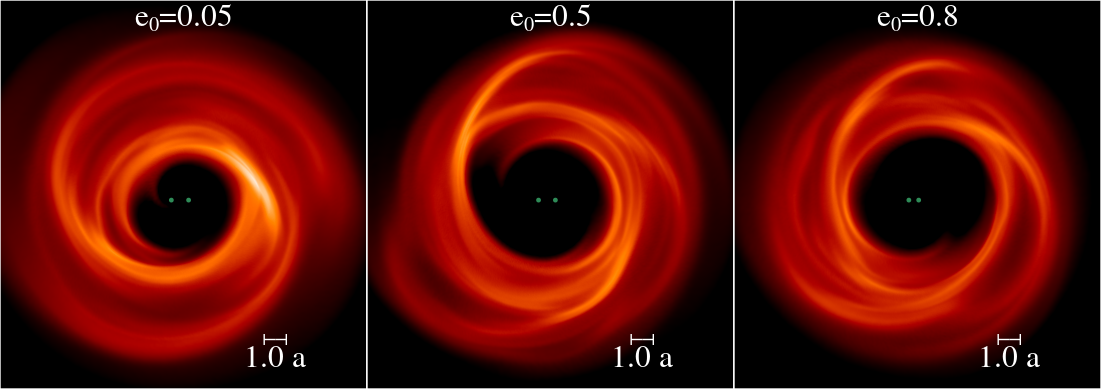}
    \caption{Column density plots of the self-gravitating circumbinary disc around the binary (shown by the green circles) at $t=320\,P_{\rm b}$.  The view is of the $x$-$y$ plane (i.e. the binary orbital plane) and the density has been integrated through $z$. The logarithmic color scale spans about two orders of magnitude in density and is the same for all the plots.}
    \label{fig:qecc2}
\end{figure}

The outward transport of angular momentum that the disc extracts from the binary, ultimately causes a radial expansion of the disc and therefore weakens the interaction with the binary. This is more evident in very eccentric systems, where the cavity carved by the binary is larger. Figure \ref{fig:qecc2} shows the morphology of the discs around an almost equal mass binary ($q=0.9$) but with different initial eccentricities.  
It is important to notice, however, that this is unlikely to be the main driver of eccentricity saturation. This is evident by looking at the $e/e_0$ versus $a/a_0$ evolution, shown in the right panel of Figure \ref{fig:aeevo}. The slope of this curve is much steeper from the $e_0=0.05$ systems than for the higher eccentricity ones. This means that for these latter systems, the interaction with the disc is still efficient enough to affect the binary, causing a significant orbital shrinking. Despite this, the eccentricity hardly changes, indicating that saturation is being achieved. It is still possible that the eccentricity will evolve over tens of thousands of orbits, but it is clear that both systems with $e_0=0.5, 0.8$ have reached a quasi-equilibrium eccentricity, which is different for the two systems. 

These results point towards the existence of a {\it range} of critical eccentricities, rather than a single value. For an almost equal mass binary with $q=0.9$, this range seems to be $e_{\rm crit}\in[0.4, 0.7]$, with initially highly eccentric binaries reaching higher values of $e_{\rm crit}$.

The case $q=0.3$ (middle panels of Fig. \ref{fig:aeevo}) displays essentially the very same features. The $e_0=0.8$ binary saturates at $e_{\rm crit}\approx 0.7$; the $e_0=0.5$ binary only shows a mild eccentricity growth possibly saturating around $e_{\rm crit}\approx 0.5$; the $e_0=0.05$ system is still evolving although the eccentricity growth rate is slowing down indicating a possible saturation around $e_{\rm crit}\approx 0.4$. 

The case $q=0.1$ is a bit different, mostly because the initially very eccentric binary experiences a violent transient that brings its eccentricity down to $\approx 0.5$ after 100 orbits. In this case, it appears that the range of $e_{\rm crit}$ is narrower, i.e. $e_{\rm crit}\in[0.4, 0.5]$. It is, however hard to estimate how much this depends on the spurious transient in the first 100 orbits of the simulation. It is also possible that for such small $q$ a well defined  $e_{\rm crit}$ exists. In fact, the upper right panel of Figure \ref{fig:aeevo} shows that the $e$ vs $a$ evolution is progressively faster for less eccentric binaries. This indicates that, while the $e_0=0.8$ system has reached a steady configuration, the other two systems are still evolving towards higher eccentricities. 

\subsection{Cooling}

\begin{figure}
    \centering
    \includegraphics[width=\columnwidth]{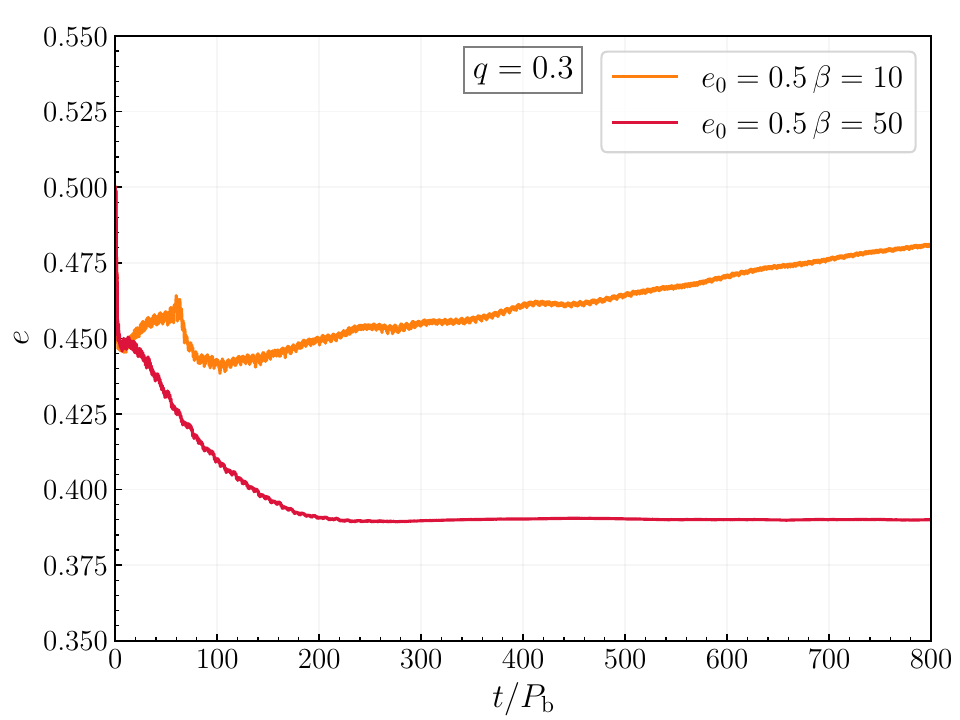}
    \caption{Evolution of eccentricity with initial $e_0=0.5, q=0.3$ for $\beta=10$ (orange line) and $\beta=50$ (red line). }
    \label{fig:cooling}
\end{figure}

The development of GIs that transport angular momentum inside the disc depends on the cooling timescale. In order to investigate how different cooling timescales affect our results in terms of binary eccentricity evolution we ran the simulation with $q=0.3$ and initial eccentricity $e_0=0.5$ also with $\beta=50$, i.e. assuming the cooling to occur on a much longer timescale. Note that $\sim 1000$ binary orbits in this case corresponds to $\sim 2$ cooling times at the disc outer edge.
The evolution of the binary eccentricity is shown in Figure \ref{fig:cooling} where the orange line shows the $\beta=10$ case while the red line shows the $\beta=50$ case.

\begin{figure}
    \centering
    \includegraphics[width=\columnwidth]{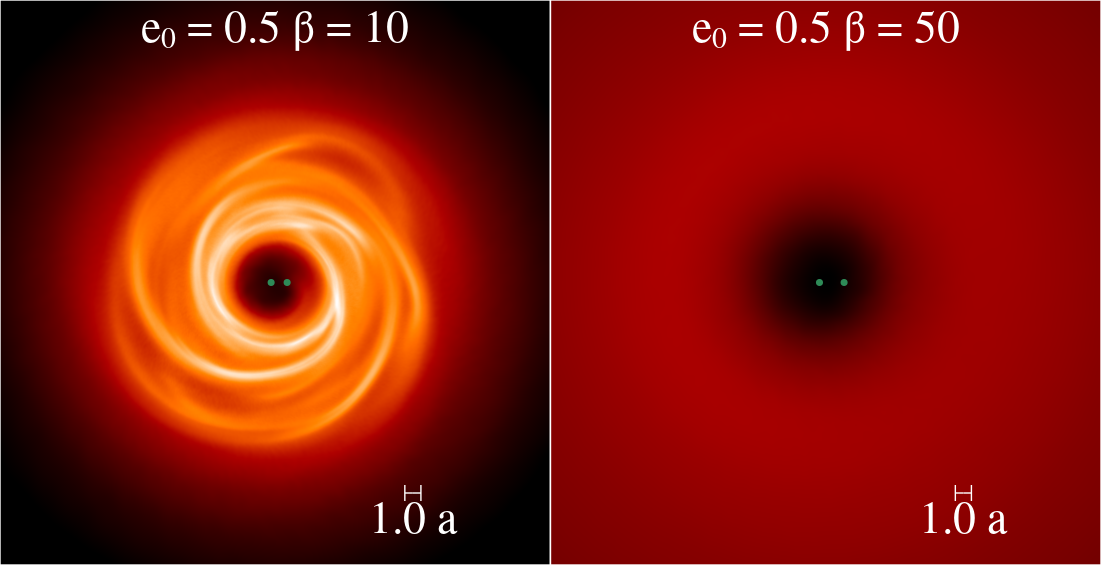}
    \caption{Column density plots of the self-gravitating circumbinary disc around the binary (shown by the green circles).  The view is of the $x$-$y$ plane (i.e. the binary orbital plane) and the density has been integrated through $z$. The logarithmic color scale spans about two orders of magnitude in density and is the same for all the plots. }
    \label{fig:cooling2}
\end{figure}

The cooling timescale is much longer than the dynamical time and therefore the disc does not develop GIs that transport the angular momentum outwards within the 900 orbits we simulated. Figure \ref{fig:cooling2} shows the column density plots of the two circumbinary discs. The left panel shows the GIs in the disc while the right panel does not show any spiral since the disc is warmer and more stable against cooling.

A longer cooling timescale ultimately results in a lower final binary eccentricity.
After the initial spurious transient, the eccentricity in the $\beta=50$ case (red line in Fig. \ref{fig:cooling}) remains essentially constant at around $e=0.38$. 

The disc stability parameter reaches a value $Q\approx20$ in the $\beta=50$ case while it remains between 1 and 2 in the fiducial case with a faster cooling timescale.
Similarly, the disc aspect ratio reaches $H/R\sim 0.3$ in the slow-cooling case while it stabilizes around $H/R\sim 0.05$ in the fiducial case, consistently with the results of \cite{franchini2021}.

\subsection{Measuring viscosity induced by GIs}

It is interesting to measure the level of viscosity in our discs with self-gravity, which essentially indicates how effective GIs are in transporting angular momentum.
The main reason is that the evolution of the binary orbital properties critically depend on the efficiency with which the angular momentum is transferred from the binary to the disc.
We can use the so-called 'GI Wiggles' \citep{hall2020,longarini2021,Terry2022}, which are essentially deviations from the Keplerian velocity field generated by the spiral density wave, to infer the $\alpha$ induced by GIs within the disc. In particular, we use the radial velocity perturbation to constrain $\alpha$ in the simulation. According to \cite{longarini2021}, the amplitude of the radial velocity perturbation is 
\begin{equation}
    \mathcal{A}=-3m\alpha^{1/2}\left(\frac{M_{\rm d}}{M_\star}\right)^2 v_{\rm K}(R),
\end{equation}
where $m$ is the number of spiral arms, set by inspecting the density structure of the simulations, and $v_{\rm K}$ is the Keplerian velocity.
We apply this method to the simulations with binary with mass ratio $q=0.3$, initial eccentricity $e_0=0.5$ and for the two different cooling parameters $\beta=10$ and $\beta=50$.
The corresponding value of the viscosity parameter is $\alpha=0.05$ for the fast cooling while it is $\alpha=0.008$ for the slow cooling rate. This is perfectly consistent with the choice of $\beta$ values since discs with higher cooling rate (i.e. smaller $\beta$) promptly form spiral structures that transport angular momentum.

The lower value of the disc viscosity in the simulation with $\beta=50$ is possibly the reason why the binary eccentricity does not evolve significantly after the initial transient phase.
We note here that while in the $\beta=10$ case the binary semi-major axis simply decreases with time, in the $\beta=50$ simulation the binary experiences an initial phase of expansion, lasting the first 200 orbits, before starting to shrink. This is somewhat expected as the initial disc aspect ratio is very large (i.e. $H/R=0.2$) and comparable to the disc mass normalized by the binary mass. Therefore the disc starts with $Q \sim 1$ and the inefficient cooling allows its temperature, thus its aspect ratio, to increase leading to binary expansion \citep{tiede2020,heathnixon2020,Franchini2022}.
However, after the first 200 orbits, the disc has reached a new thermal equilibrium, as its mass has significantly decreased, allowing the binary to shrink.

We also measured the viscosity parameter comparing simulations with different mass ratio and eccentricity but with the same $\beta=10$. We find the initial eccentricity value not to dramatically change the value of $\alpha$. Comparing binaries with $e_0 = 0.5$ and $e_0 = 0.8$ (same mass ratio $q=0.3$) we find $\alpha$ to be $14\%$ larger for less eccentric binaries.
Instead, higher binary mass ratios cause a $30\%$ larger viscosity compared to more equal mass binaries. We measure the viscosity of the disc surrounding the binary with $q=0.1$ and $q=0.3$ to be $\alpha=0.0726$ and $\alpha=0.05$ respectively.

\subsection{Accretion rate periodicity}

We also investigated the periodicity of the accretion rate onto the binary. This is expected to be modulated on the binary and disc dynamical time scales \citep{roedig2011}. 

\begin{figure}
    \centering
    \includegraphics[width=\columnwidth]{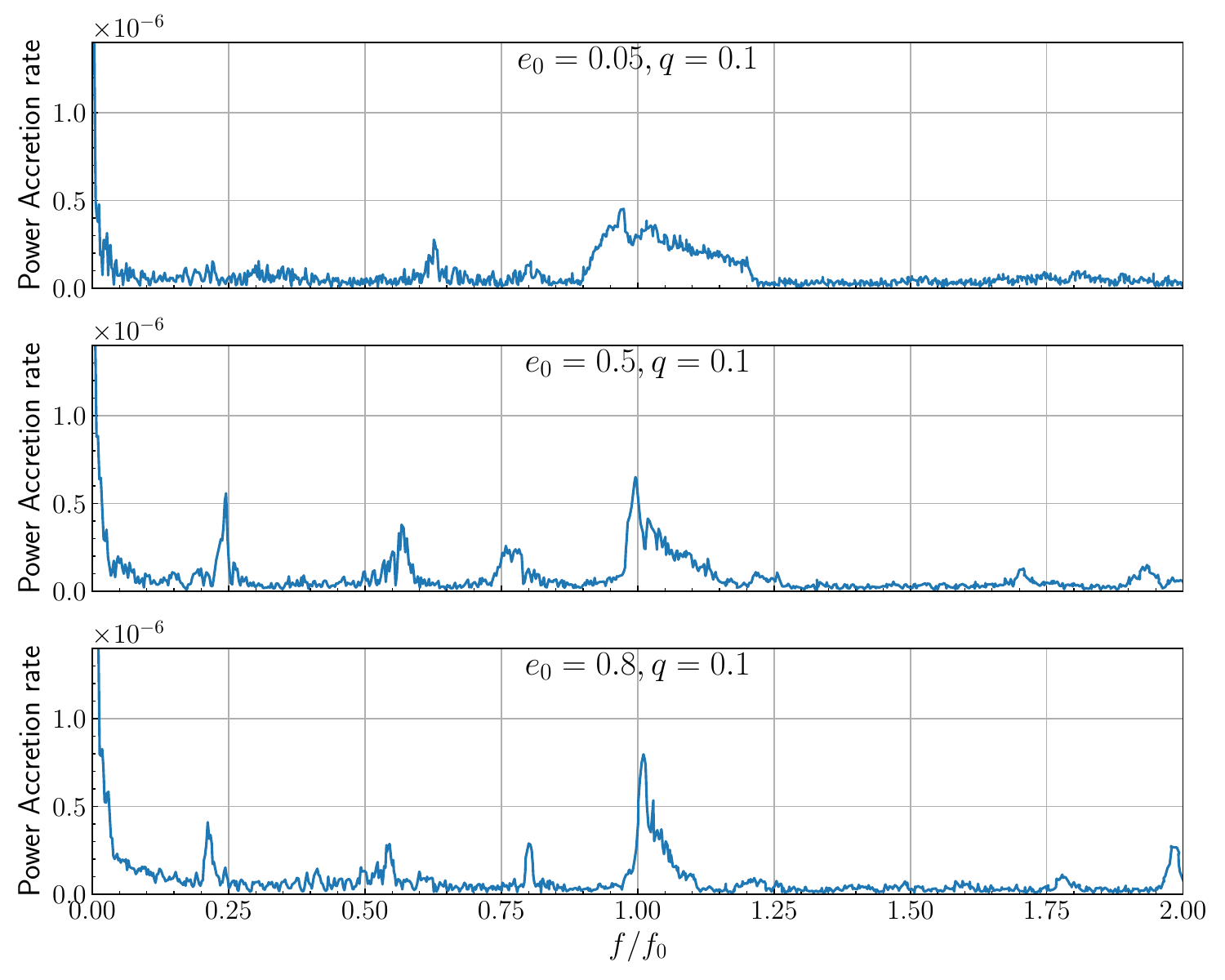}
    \includegraphics[width=\columnwidth]{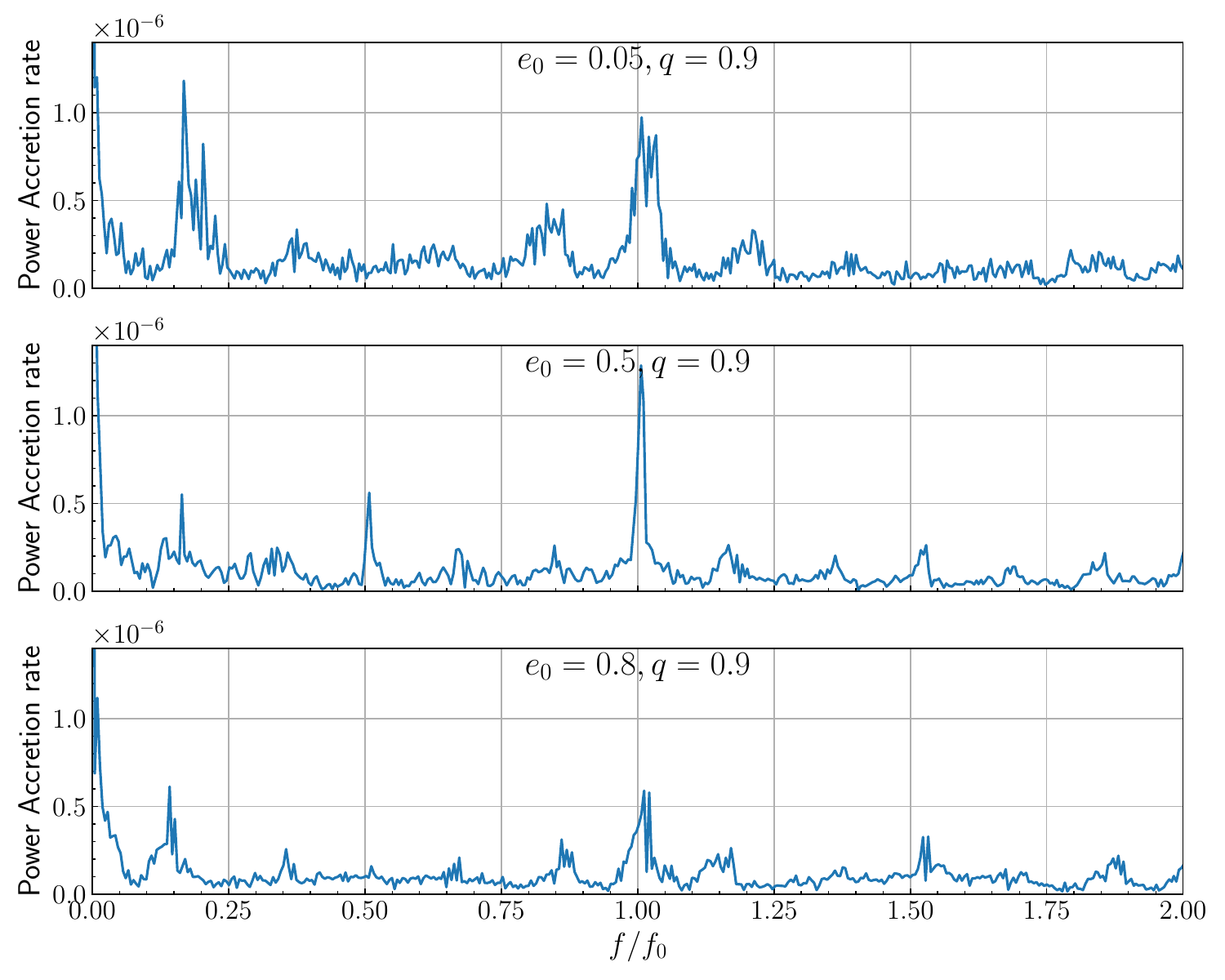}
    \caption{FFT of the accretion rate onto the binary for $e_0=0.05$ (upper panel), $e_0=0.5$ (middle panel) and $e_0 = 0.8$ (lower panel). The first three panels are for binaries with initial mass ratio $q=0.1$ while the last three have $q=0.9$.}
    \label{fig:ffts}
\end{figure}

We computed the Fast Fourier Transform (FFT) of the accretion rate in our simulations with different initial binary parameters. The results are shown in Figure \ref{fig:ffts} where the frequency is in units of the initial binary orbital frequency $f_0=2\pi(GM/a^3)^{1/2}$.
We find the peaks that correspond to the binary, i.e. $f = f_0$ and disc, i.e. $f\simeq 0.2f_0$ periodicities to become more narrow with increasing eccentricity for unequal mass binaries. 
In very unequal mass, almost circular binaries we find the binary frequency peak to be significantly broader, indicating that the binary orbit does evolve considerably throughout the simulation. Furthermore, the periodicity associated with the disc inner edge is swamped by the sampling noise of the accretion rate. 
This is consistent with previous works \citep{farris2014} that find the disc frequency to be prominent only for equal mass binaries, which efficiently excite an $m=1$ mode, also called 'lump', at the disc cavity edge.
We do indeed recover the lump periodicity in the almost equal mass case as shown in the bottom three panels of Figure \ref{fig:ffts}.
The presence of the lump is also associated with binary eccentricity. We indeed find this modulation to be prominent even in eccentric, unequal mass binaries.
Note that for both mass ratios shown in Figure \ref{fig:ffts}, the lump peak slightly shifts at lower frequencies for larger eccentricities, consistent with a larger cavity (see Fig. \ref{fig:qecc2}). Furthermore, the peak is slightly shifted to lower frequencies for almost equal mass binaries, still consistent with a larger cavity. 

The amplitude of the binary orbit modulation tends to increase with increasing mass ratio for binaries with low to mild eccentricities, i.e. $e\sim 0.05-0.5$, while it remains roughly constant for very eccentric binaries (see third and sixth panels of Fig. \ref{fig:ffts}).

Our results in terms of periodicities associated with MBHBs embedded in discs are broadly consistent with the literature \citep{dorazio2013,Munoz2020}, except we do find the modulation associated with the binary orbital period in all cases, possibly because we do evolve the binary orbit  \citep{Franchini2023}.
Although binaries with such massive discs are expected to have long periods, if we extrapolate these findings to more compact systems, our results would indicate that more eccentric and unequal mass  binaries should be more easily identified in photometric searches for MBHBs as the amplitude of the binary orbital period modulation is in general higher.

\subsection{Preferential accretion}

\begin{figure}
    \centering
    \includegraphics[width=\columnwidth]{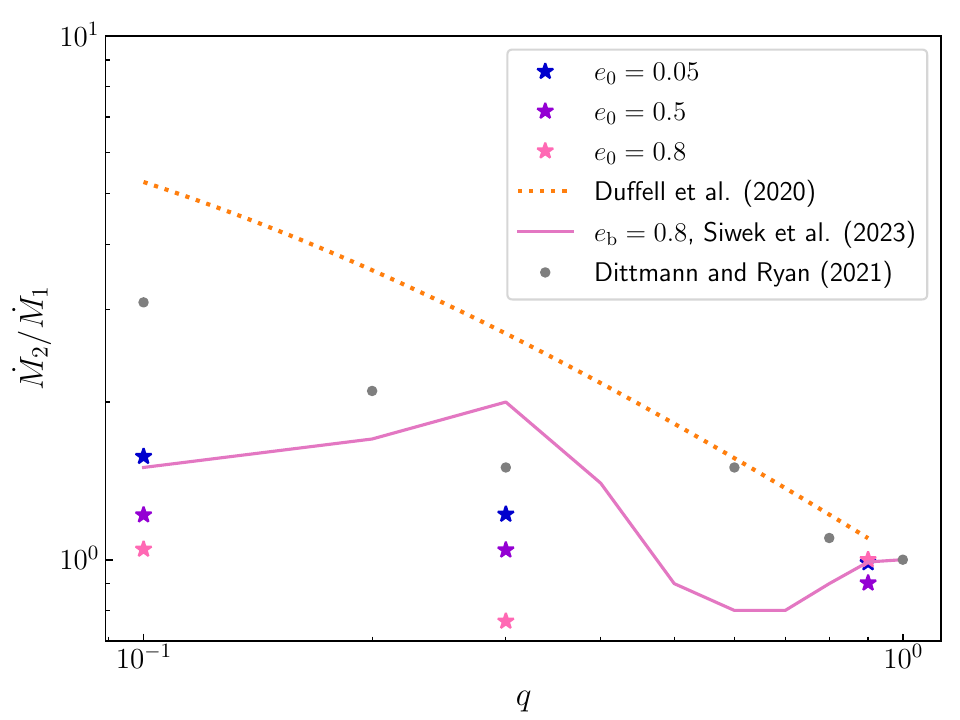}
    \caption{Mass ratio growth as a function of the initial binary mass ratio calculated over the simulated 1000$P_{\rm b}$. The blue, violet and pink stars show our results for $e_0=0.05$, $e_0=0.5$ and $e_0=0.8$ respectively. The orange dotted line shows the result obtained for circular binaries by \cite{Duffell2019} while the pink solid line shows the result obtained for binaries with $e_{\rm b}=0.8$ by  \cite{Siwek2022}. The grey dots show the values obtained by \cite{Dittmann2021} for circular binaries. }
    \label{fig:pref_acc}
\end{figure}

Finally, we investigated the preferential accretion of gas onto the secondary MBH as a function of initial binary eccentricity and mass ratio.
We calculated the accretion rate onto each binary component over the entire 1000 orbits we simulated. 
Figure \ref{fig:pref_acc} shows the ratio between the accretion rate on the secondary over the primary as a function of the initial binary mass ratio for different values of the initial binary eccentricity.
We also show the results obtained for circular binaries by  \cite{Duffell2019} (orange dotted line) and \cite{Dittmann2021} (grey dots) and for highly eccentric binaries by \cite{Siwek2022} (i.e. the pink line in their Fig. 4).

We find initially less eccentric binaries to experience a stronger preferential accretion onto the secondary, which is consistent with previous studies in the same self-gravitating regime \citep{roedig2011}.
A similar trend has been found in previous numerical works employing 2D hydrodynamics simulations \citep{Duffell2019}. 
However, this trend is much milder in our case and consistent with \citep{roedig2011}.
We do however note that the results presented in \cite{Dittmann2021} for circular binaries show a milder trend with mass ratio compared to other 2D simulations works.

Note that there is still a strong preferential accretion on the secondary in terms of Eddington ratio. Indeed, $\dot{M}_2/\dot{M}_1=1.5$ for $q=0.1$ implies that the secondary is accreting at a 15 times higher Eddington ratio, so in this sense there is still strong preferential accretion on the secondary. 

We find that mass ratios grow more efficiently in low to moderately eccentric binaries while the growth is suppressed for highly eccentric binaries.
We can see from Figure \ref{fig:pref_acc} that for fixed binary mass ratio, the mass ratio growth is suppressed as the eccentricity increases to the point where the primary is accreting more gas than the secondary, which occurs for $q=0.3$ and $e_0=0.8$.
This is broadly consistent with the pink line in Figure 4 in \cite{Siwek2022}, although they find preferential accretion on the primary for slightly higher mass ratios compared to $q=0.3$.

We note that the relation for preferential accretion as a function of mass ratio inferred in previous studies, e.g. \cite{Munoz2020,Duffell2019,Siwek2022}, overestimates the mass ratio growth for low mass ratios compared with our simulations. 
A possible reason for this discrepancy might be the 3D geometry we use. We do simulate a finite disc thickness in the vertical direction, therefore the gas can more easily avoid the secondary and reach the primary instead.
The comparison with previous studies is however not straightforward  as our simulations include the gas self-gravity and allow the gas to heat and cool. The dynamics of the gas inside the cavity is therefore different compared with the locally isothermal case and this ultimately leads to a different interaction between the binary and the disc. Furthermore, the accretion of matter onto the two binary components is regulated by the GIs transporting angular momentum in our simulations. This is a major difference between our work and previous similar numerical works, as the effective viscosity provided by GIs might be significantly lower than the value assumed in previous studies. 

We note that if $\beta$ is larger (e.g. $\beta=50$), meaning less efficient cooling, then the mass ratio growth is $\dot{M}_2/\dot{M}_1=1.06$ for $q=0.3$, $e_0=0.8$. This is due to the fact that the disc remains warmer compared to the $\beta=10$ case.
Preferential accretion has also been investigated by \cite{Young2015}, who showed that an increase in the
gas temperature allows more material to accrete on to the primary component of the binary. The relative accretion by the primary was also found to increase at least quadratically with the sound speed of the accreted gas. Since however in our case all discs settle to roughly the same temperature as a result of self-regulation (note that they all have the same initial mass and aspect ratio), we do not speculate on the dependence of our results on the disc temperature.

\section{Conclusions}
\label{sec:conclusions}

In this work, we explored the dynamics of sub-pc MBHBs interacting with a massive circumbinary gaseous disc whose self-gravity is not negligible compared to the binary gravitational potential.
We ran a suite of numerical simulations using the SPH code {\sc phantom}, changing the initial binary eccentricity $e_0$, mass ratio $q$ and cooling factor $\beta$. We ran all the simulations for 1000 binary orbits, in order to ensure that all systems reached a quasi-steady state configuration.

The main aim of this work was to study the eccentricity evolution in these binary systems in order to search for a limiting value, which was speculated from previous simulations \citep{roedig2011}.
The main implication of the existence a critical eccentricity is that we could in principle understand whether this binary has significantly interacted with its gaseous environment.
We do not find one value of critical eccentricity valid for all the binaries we simulated. Instead we find the limit eccentricity to depend on the initial mass ratio, in particular more unequal mass binaries tend to saturate around $e\sim 0.4-0.5$ while the range for equal mass binaries is much larger, i.e. $0.3-0.7$.
We note that the limiting eccentricity value of the binary is also sensitive to the cooling choice. We measure the value of $\alpha$ that GIs generate and we indeed find slower cooling to lead to lower viscosity values which ultimately translates in the binary stalling at a lower value of eccentricity compared to faster cooling rates.

Our findings are in some contrast with those reported in \cite{Zrake2021}. We caution however that the direct comparison is not completely fair as we employ 3D hydrodynamical simulations of live binaries in a region of the parameter space where the gas self-gravity cannot be neglected while \cite{Zrake2021} inferred their results for a locally isothermal circumbinary disc neglecting the gas self-gravity.
The inclusion of the gas self-gravity significantly changes the dynamics of the gas within the disc, the amount of material that is deposited at the resonances location and therefore ultimately the evolution of the binary orbital parameters.

As we have a finite gas reservoir, the interaction between the binary and the disc is expected to weaken with time. However, after 1000 orbits there is still a significant fraction of the disc, i.e. $\sim80$\%, that interacts with the binary. Furthermore, at least in the highly eccentric case, we can see that as the binary mass ratio tends towards unity the cavity becomes large (see e.g. Fig. \ref{fig:qecc2}) causing the interaction with the disc to become less strong. 

Since we simulate live binaries, we can directly trace the accretion rate onto each BH. We therefore computed the FFT of the accretion rate in order to investigate the effect of the initial binary eccentricity and mass ratio on the amplitude and frequency of the modulation.
We recover the modulation on the binary orbital period in every simulation, regardless of the initial mass ratio and eccentricity. 
The amplitude of this modulation increases with increasing mass ratio and eccentricity.
The modulation on the inner disc edge orbital period is absent in almost circular, unequal mass binaries. In this case, the cavity does not develop significant eccentricity and does not therefore precess, causing the modulation.

We also measured the binary preferential accretion rate for different eccentricities and mass ratios.
We showed that in general the secondary BH accretes more mass than the primary in low mass ratio binaries. 
The mass ratio growth is not strictly monotonic with either the mass ratio or the binary eccentricity. This is somewhat expected since the binary naturally evolves its properties during the simulation and therefore relating the preferential accretion to the initial properties of the binary might not be straightforward.
The general trend is that the mass ratio growth is suppressed at higher eccentricities and larger mass ratios, broadly consistent with previous works \citep{Munoz2020,Siwek2022}.

We have considered binaries embedded in self-gravitating discs, therefore in a regime that is relevant for PTA.
The amplitude of the GWB that is expected to be observed with PTA depends on the chirp mass of the sources with $\mathcal{M}^{5/3}$, which in turn depends on the mass ratio as $\mathcal{M}=(q/(1+q)^2)^{3/5}\,M$. The higher the mass ratio, the larger the chirp mass therefore the louder the GWB signal.
Determining the evolution of MBHB mass ratio is therefore crucial for the estimate of the signal amplitude that the population will produce. 
We find a non-monotonic dependence of the mass ratio growth on the eccentricity. In particular in the case of mass ratios close to $q=0.3$ we find an inversion of the preferential accretion for high eccentricities (i.e. $e>0.7$), meaning that the primary accretes more material and therefore the mass ratio decretes.

Note that there are no evident circumBH discs in our simulations. This is simply due to the fact that since the disc temperature regulates at a correspondent aspect ratio $H/R\sim0.03-0.04$, the amount of material that leaks into the cavity is very small, therefore its contribution to the binary evolution is negligible in this regime. The detailed investigation of the dynamics within the cavity in this particular regime is deferred to a future work.

\begin{acknowledgements}
We thank Daniel Price for providing the {\sc phantom} code for numerical simulations and acknowledge the use of {\sc splash} \citep{Price2007} for the rendering of the figures.
AF and AS acknowledge financial support provided under the European Union’s H2020 ERC Consolidator Grant ``Binary Massive Black Hole Astrophysics" (B Massive, Grant Agreement: 818691). CL acknowledges financial support by UK Science and Technology research Council (STFC) via the consolidated grant ST/W000997/1. 
\end{acknowledgements}

%
%

\bibliographystyle{aa} 
\bibliography{literature}

\end{document}